\documentstyle[prb,aps]{revtex}

\begin{document}

\twocolumn[{
\draft
\widetext
\title{Dynamical analysis of the buildup process near resonance}
\author{Jorge Villavicencio\cite{byline}}
\address{Centro de Investigaci\'on Cient\'{\i}fica y de Educaci\'on\\
Superior de Ensenada\\
Apartado Postal 2732, 22800 Ensenada, Baja California, M\'exico}
\author{Roberto Romo}
\address{Facultad de Ciencias,\\
Universidad Aut\'onoma de Baja California\\
Apartado Postal 1880, 22800 Ensenada, Baja California, M\'exico}
\date{\today}
\mediumtext

\begin{abstract}
The time evolution of the buildup process inside a double-barrier system for
off-resonance incidence energies is studied by considering the analytic
solution of the time dependent Schr\"{o}dinger equation with cutoff plane
wave initial conditions. We show that the buildup process exhibits
invariances under arbitrary changes on the system parameters, which can be
successfully described by a simple and easy-to-use one-level formula. We
find that the buildup of the off-resonant probability density is
characterized by an oscillatory pattern modulated by the resonant case which
governs the duration of the transient regime. This is evidence that
off-resonant and resonant tunneling are two correlated processes, whose
transient regime is characterized by the same transient time constant of two
lifetimes.
\end{abstract}

\maketitle
}] \narrowtext



\narrowtext

The buildup process of electrons inside the quantum well of a double barrier
(DB) resonant structure has been one of the most important problems under
investigation since it governs the ultimate speed of high frequency
tunneling devices \cite{Sollner83,Luryi}. Although the first theoretical
efforts to estimate the relevant time scales for this mechanism were based
on stationary approaches \cite{Luryi,Ricazbel}, it has been widely
recognized that this process is of a dynamical nature \cite{Ricazbel,todos};
hence, the solution of the time-dependent Schr\"{o}dinger equation provides
the most reliable way to tackle this fundamental problem.

In this letter we provide a full quantum dynamical study of the buildup
process at resonance and off-resonance incidence energies, based on an exact
analytical solution. We show that within the complexity of the process,
underlying invariances can be found in the time evolution of the probability
density. Important conclusions on the relevant time scale that characterizes
both the resonant and off-resonant buildup are brought out from such
invariances.

Our analysis deals with the analytic solution of the time-dependent
Schr\"{o}dinger equation for a finite range potential $V(x)$ that vanishes
outside the region $0\leq x\leq L$, with a cutoff plane initial condition, 
\begin{equation}
\Psi \left( x,k;t=0\right) =\left\{ 
\begin{array}{cc}
e^{ikx}-e^{-ikx}, & \quad -\infty <x\leq 0, \\ 
0, & \quad x>0,
\end{array}
\right.  \label{obtur}
\end{equation}
which refers to a perfect reflecting shutter \cite{Note1}. The solution for
the internal region \cite{Note2} reads, 
\begin{eqnarray}
\Psi (x,k;t) &=&\phi (x,k)M(0,k;t)-\phi ^{\ast }(x,k)M(0,-k;t)  \nonumber \\
&&-i\sum\limits_{n=-\infty }^{\infty }\phi _{n}M(0,k_{n};t),  \label{Psiint}
\end{eqnarray}
where $\phi (x)$ stands for the stationary solution, and $\phi
_{n}=2ku_{n}(0)u_{n}(x)/(k^{2}-k_{n}^{2})$ is given in terms of the resonant
states, $\{u_{n}(x)\}$, and the $S-$matrix poles, $\{k_{n}\},$ of the
system. The index $n$ runs over the complex poles $k_{n}$, distributed in
the third and fourth quadrants in the complex $k$-plane. In the above
equation the Moshinsky functions $M(0,q;t)$ are defined in terms of the
complex error function $w(z)$, as $M(0,q;t)\equiv M(y_{q})=w(iy_{q})/2$,
where the argument is given by $y_{q}=-exp(-i\pi /4)(m/2\hbar )^{1/2}\left[
(\hbar q/m)t^{1/2}\right] $, and $q$ stands either for $\pm k$ or $k_{\pm n}$%
.

The formal solution for the internal region, Eq. (\ref{Psiint}), allows us
to study both the spatial behavior and the time evolution of the probability
density for any incidence energy{\em \ }$E=\hbar ^2k^2/2m$.{\em \ }The main
ingredients of Eq. (\ref{Psiint}) are the stationary wavefunction $\phi
(x,k) $, the resonance parameters{\em \ }\{$E_n=\varepsilon _n-i\Gamma
_n/2=\hbar ^2k_n^2/2m$\}{\em \ }and the corresponding resonant
eigenfunctions $\{u_n(x)\}$. The latter can be obtained by a straightforward
calculation using the transfer matrix method adapted to the complex
eigenvalue problem \cite{GCRRAR}. For systems with isolated
(non-overlapping) resonances, the single-resonance term approximation to Eq.
(\ref{Psiint}) gives an excellent description, except for very short times ($%
t\ll \hbar /\Gamma _n$) in which one has to consider contributions from
additional resonance terms. In all the numerical examples presented here the
single term approximation applies.

As a first example, let us consider the DB structure (system A) with
parameters: barrier heights $V_1=V_2=0.23$ $eV$, barrier widths $b_1=b_2=5.0$
$nm$, well width $\omega _0=5.0$ $nm$, and effective mass for the electron $%
m=0.067m_e$. The resonance parameters for the first resonant state are:
energy position, $\varepsilon _1=80.11$ $meV$, and resonance width, $\Gamma
_1=1.03$ $meV$. In Fig. \ref{fig1} (a) we plot $|\Psi |^2$, calculated from
Eq. (\ref{Psiint}), as a function of the position $x$ along the internal
region,{\em \ }for specific times $\{t_i\}{\em \ }$whose increasing values
are given in the figure. Here, the incidence energy is chosen below
resonance, at $E=75.0$ $meV$. Note that the off-resonant buildup occurs in
such a way that $|\Psi |^2$ is found sometimes above or below the asymptotic
value $|\phi |^2$.${\em \ }$This behavior is dramatically different from the
monotonic growth that characterizes the special case of incidence at
resonance, see Fig. 4 of Ref. 8. In order to show the time dependence of the
probability density for a fixed position $x_0$, we plot in part (b) $|\Psi
(x_0,k;t)|^2$\ versus $t$ for different deviations from resonance $\Delta
E=|E-\varepsilon _1|$. Note that $|\Psi |^2$\ fluctuates around its
asymptotic value $|\phi |^2$, and exhibits an oscillatory behavior not
present in the resonant case, as shown in Fig. \ref{fig1} (b).

Up to here, we have illustrated the behavior of the off-resonance buildup
only for a particular potential profile. It is clear that any changes in
either the incident energy or the potential profile parameters will affect
the solution $\Psi (x,k;t)$ since the relevant input to Eq. (\ref{Psiint}),
namely, $\phi \left( x,k\right) $, $u_n\left( x\right) $ and $E_n$, strongly
depends on the potential parameters. For instance, let us consider two
additional DB systems with potential profiles quite different from system A;
the first corresponds to the symmetrical structure (system B) with parameters:
barrier heights $V_1=V_2=0.5$ $eV$, barrier widths $b_1=b_2=3.0$ $nm$\ and
well width $\omega _0=10.0$ $nm$; the second corresponds to an asymmetrical
structure (system C), whose parameters are: $V_1=0.45$ $eV,$ $V_2=0.35$ $eV$%
, $b_1=3.0$ $nm$, $b_2=10.0$ $nm$ and $\omega _0=8.0$ $nm$. The resonance
parameters for the first eigenstate are: $\varepsilon _1=37.80$ $meV$, $%
\Gamma _1=0.12$ $meV$ (system B); and $\varepsilon _1=51.29$ $meV$, $\Gamma
_1=0.17$ $meV$ (system C). The value of the transmission peak $T(\varepsilon
_1)$ is unity for A and B; for system C, $T(\varepsilon _1)<1$, since it is
asymmetric. We choose here the incidence energies $E=\varepsilon _1-\Delta E$%
\ such that the ratio $\gamma =T(E)/T(\varepsilon _1)$\ is the same for A,
B, and C. For example, if we choose $E$ such that $T(E)$ is $1.0$ $\%$ of $%
T(\varepsilon _1)$, from numerical inspection from a $T$ versus $E$ plot
(not shown here), the incidence energies for A, B, and C, must be $74.97$, $%
37.20$ and $50.44$ ($meV$), respectively. The results of the comparison of
the time evolution of $|\Psi |^2$ for this selection ($\gamma =.01$), are
shown in Fig. \ref{fig2} (a). The three curves are strongly different, as
expected. Thus, the complete characterization of the buildup process for a
broad range of potential geometries seems to be a too involved task; however,
one of the purposes of this work is to show that despite this complex
situation, the probability density has striking invariances under changes in
the potential profiles. To illustrate the above let us consider a more
suitable representation for the probability density {\it i. e.} $|\Psi /\phi
|^2$\ as a function of $\tau $, which is the time normalized to lifetime
units. We find a striking result: all curves coincide exactly for the three
different systems, see Fig. \ref{fig2} (b). This result suggests the
existence of an underlying invariance in the process.

In order to show the existence of such invariance, we derive a one-level
formula for the normalized probability density starting from the formal
solution (\ref{Psiint}). We proceed along the same lines discussed in our
recent work \cite{PRBC}, but considering incidence energies different from
resonance ($E\neq \varepsilon _n$). Following such a procedure we obtain,

\begin{equation}
\left| \Psi \left( \tau \right) /\phi \right| ^2=1+e^{-\tau }-2e^{-\tau
/2}\cos \left[ \omega _n\tau \right] ,  \label{Psicos}
\end{equation}
where $\omega _n=(\varepsilon _n-E)/\Gamma _n$ is a dimensionless frequency.
The reliability of this one-level formula is shown in a plot of Eq. (\ref
{Psicos}) included in Fig. \ref{fig2} (b), and we see an excellent
agreement. Furthermore, note that Eq. (\ref{Psicos}) depends in general on
the system parameters through the frequency $\omega _n$; however, it can be
shown straightforwardly that the condition previously imposed on the ratio $%
T(E)/T(\varepsilon _n)$ guarantees the independence of Eq. (\ref{Psicos}) on the
potential profile. Consider the Breit-Wigner expression for the transmission
coefficient \cite{GCRRAR},

\begin{equation}
T(E)=\frac{\Gamma _n^0\Gamma _n^L}{(E-\varepsilon _n)^2+\Gamma _n^2/4},
\label{B-W}
\end{equation}
where $\Gamma _n^0$ and $\Gamma _n^L$ are the partial decay widths of the
system which satisfy $\Gamma _n=\Gamma _n^0+\Gamma _n^L$. This formula of $%
T(E)$ is valid for isolated non-overlapping resonances, that is $\Gamma
_n\ll |\varepsilon _n-\varepsilon _{n\pm 1}|$, which is the case of a broad
range of typical DB structures. From Eq. (\ref{B-W}) we can easily calculate the
ratio $\gamma =T(E)/T(\varepsilon _1)$ and obtain an expression for the
frequency $\omega _1=\left( \gamma ^{-1}-1\right) ^{1/2}/2$. As a
consequence, Eq. (\ref{Psicos}) is no longer dependent on the system
parameters since it only depends on $\gamma $. Note that the frequency $%
\omega _1$ also measures the deviations of the incidence energy $E$ from the
resonance $\varepsilon _1$ in multiple numbers of $\Gamma _1$, {\it i.e.} $%
\Delta E=\omega _1\Gamma _1$. In other words, different systems share the
same curve of the probability density provided that deviations from
resonance are the same in units of the corresponding resonance width $\Gamma
_1$. Note also that Eq. (\ref{Psicos}) is independent of the choice $\pm \omega
_n$, which implies that deviations above and below resonance give the same
result. Since in our example $\gamma =0.01$, we have that $\omega _1\approx
5.0$; this can also be verified by computing the values of $\omega
_1=|\Delta E|/\Gamma _1$ from the incidence energies used in Fig. \ref{fig2}.

Note another interesting regularity of $\left| \Psi \left( \tau \right)
/\phi \right| ^2$; the damped oscillatory behavior in Eq. (\ref{Psicos}) is
modulated by the lower envelope

\begin{equation}
\left| \Psi \left( \tau \right) /\phi \right| ^2=\left( 1-e^{-\tau /\tau
_0}\right) ^2,  \label{envel}
\end{equation}
which is exactly the capacitor-like buildup law obtained for the special case of
incidence at resonance \cite{PRBC}, where the transient time constant $\tau
_0$ of the process is exactly two lifetimes, $\tau _0=2$. A plot of Eq. (\ref
{envel}) is included in Fig. \ref{fig2} (b). This result is relevant from a
physical point of view, since it is a manifestation of the subtle interplay
between the incident off-resonant carriers and the quasibound state of the
system: resonant and off-resonant buildup, although different processes, are
not uncorrelated at all, the latter is governed by the former in the way
exhibited in Fig. 2(b). As a consequence, the transient regime for both
situations is characterized by the same transient time constant $\tau _0$.%
{\em \ }The above mentioned quantity is relevant for the design and optimization of
resonant tunneling diodes; in this respect, a detailed discussion can be
found in a recent work by Luryi and Zaslavsky \cite{LurZas}, in which the
distinction between capacitive and quantum contributions to an effective
time constant is analyzed.

In conclusion, the dynamics of the buildup mechanism at off-resonance
incidence energies has been explored in typical DB resonant structures. We
have shown that, despite the complexity that characterizes the dynamical
process, the time evolution of the probability density exhibits invariances
under arbitrary changes on the system parameters. From such invariances we
conclude that the transient regime in both resonant and off-resonant
processes is characterized by the same transient time constant of two
lifetimes. Our results are valid for any DB system with isolated resonances
and incident plane wave initial condition.

The authors acknowledge financial support from Conacyt, M\'exico, through Contract
No. 431100-5-32082E. The authors also thank G. Garc\'{\i }a-Calder\'on for useful
discussions.

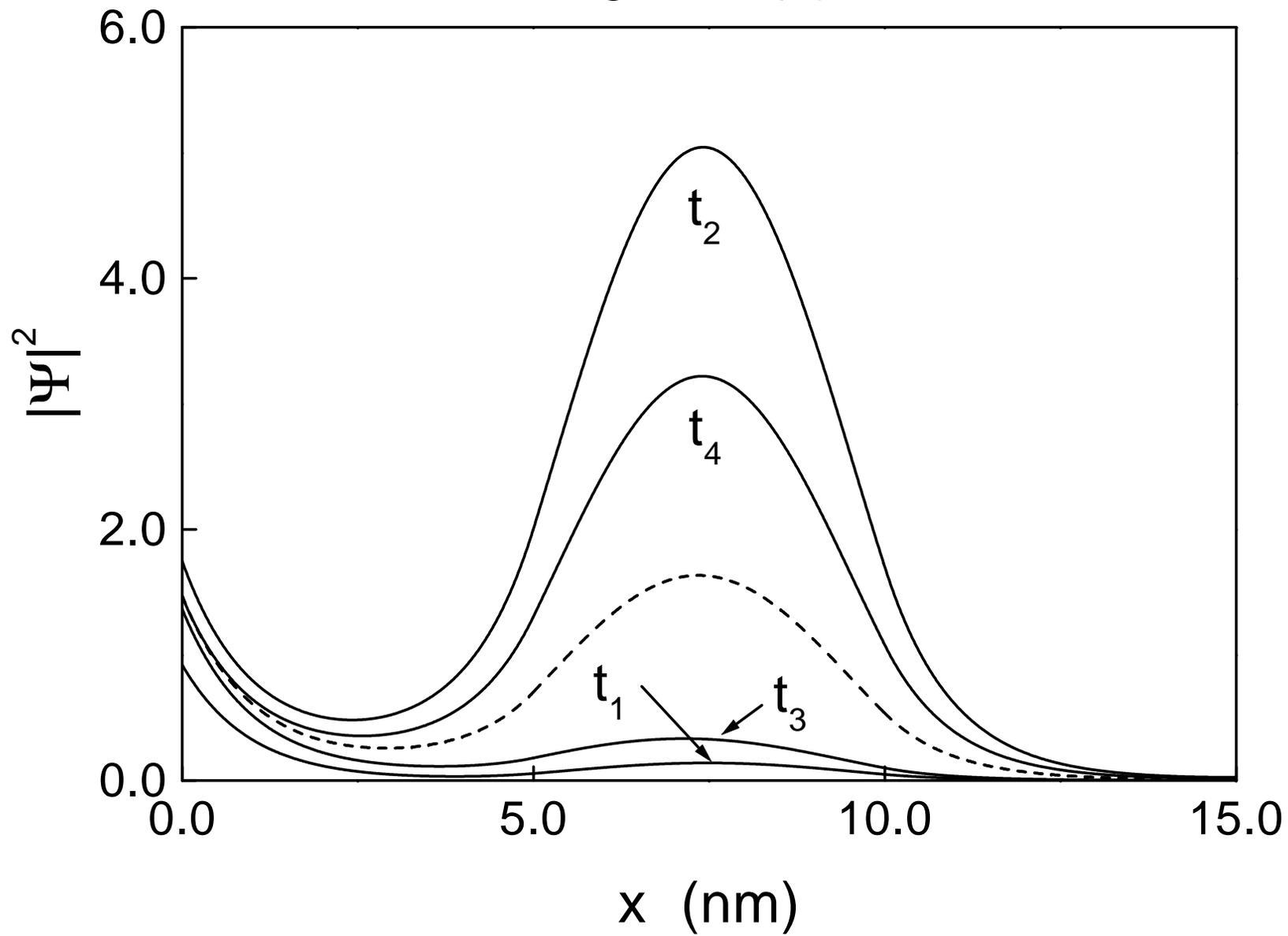
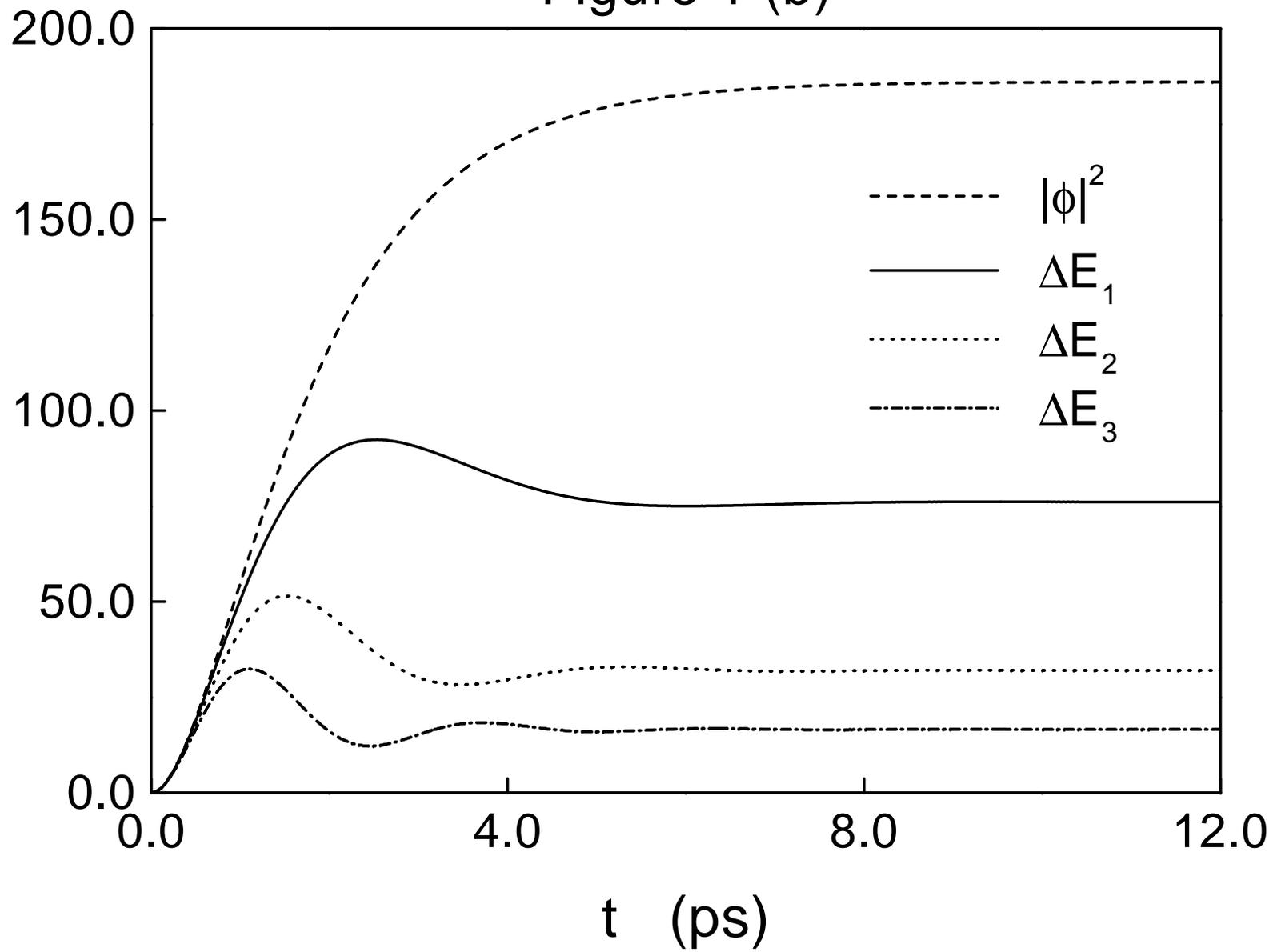
\begin{figure}[tbp]
\caption{ (a) The birth of $|\Psi |^2$ inside the
structure as a function of the position $x,$ for increasing values of time: $%
t_1=0.04$ $ps $, $t_2=0.4$ $ps$ $t_3=0.8$ $ps$, and $t_4=1.2$ $ps$ (solid
lines). The stationary solution $|\protect\phi |^2$ (dashed line) is also
included for comparison. (b) The time evolution of $|\Psi |^2$ at
the fixed position at the center of the well, for different values of $%
\Delta E = \Delta E_k$, where: $\Delta E_1 = 0.6, \Delta E_2 = 1.1$, and $%
\Delta E_3 = 1.6$ (meV).}
\label{fig1}
\end{figure}

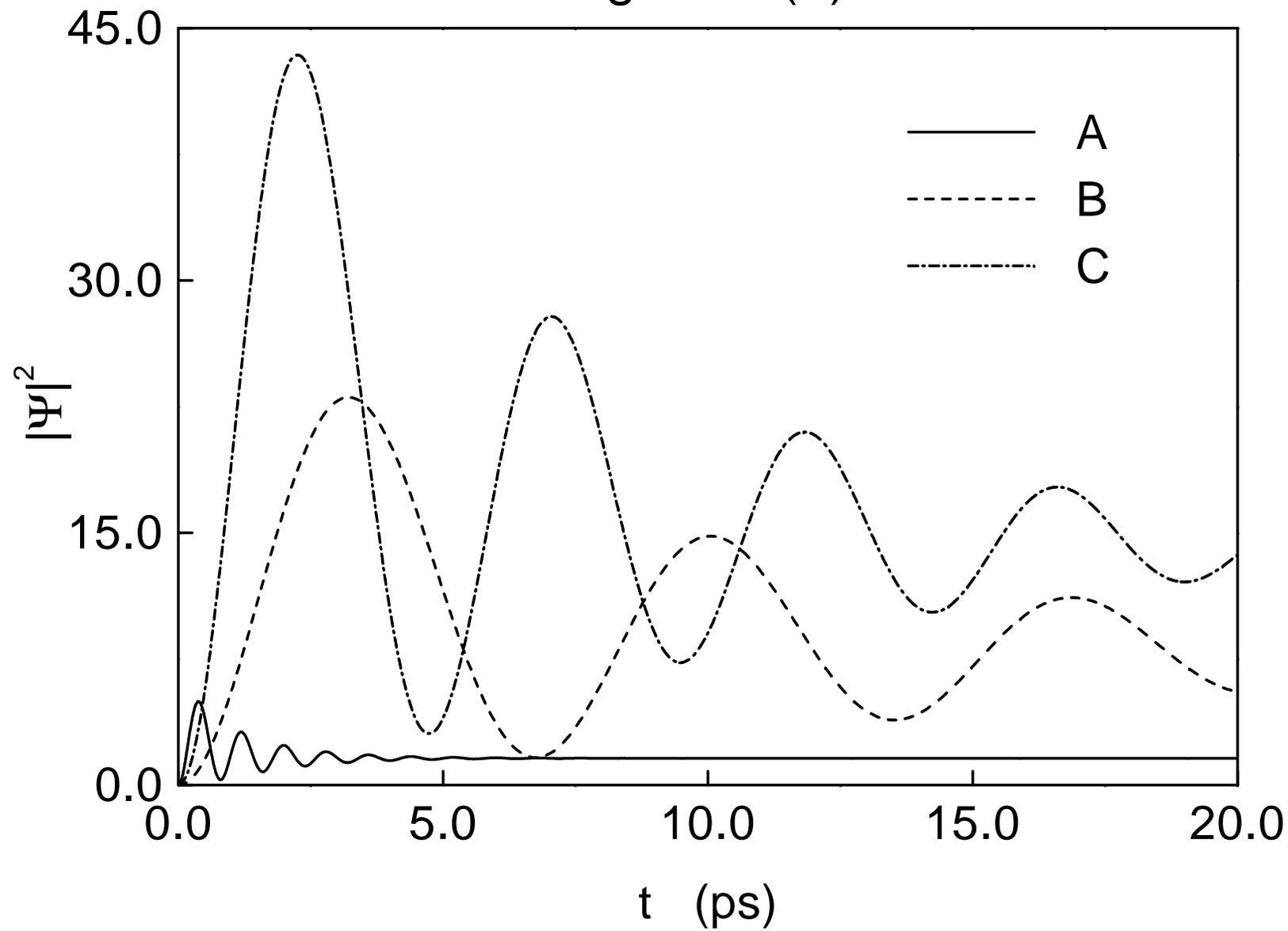
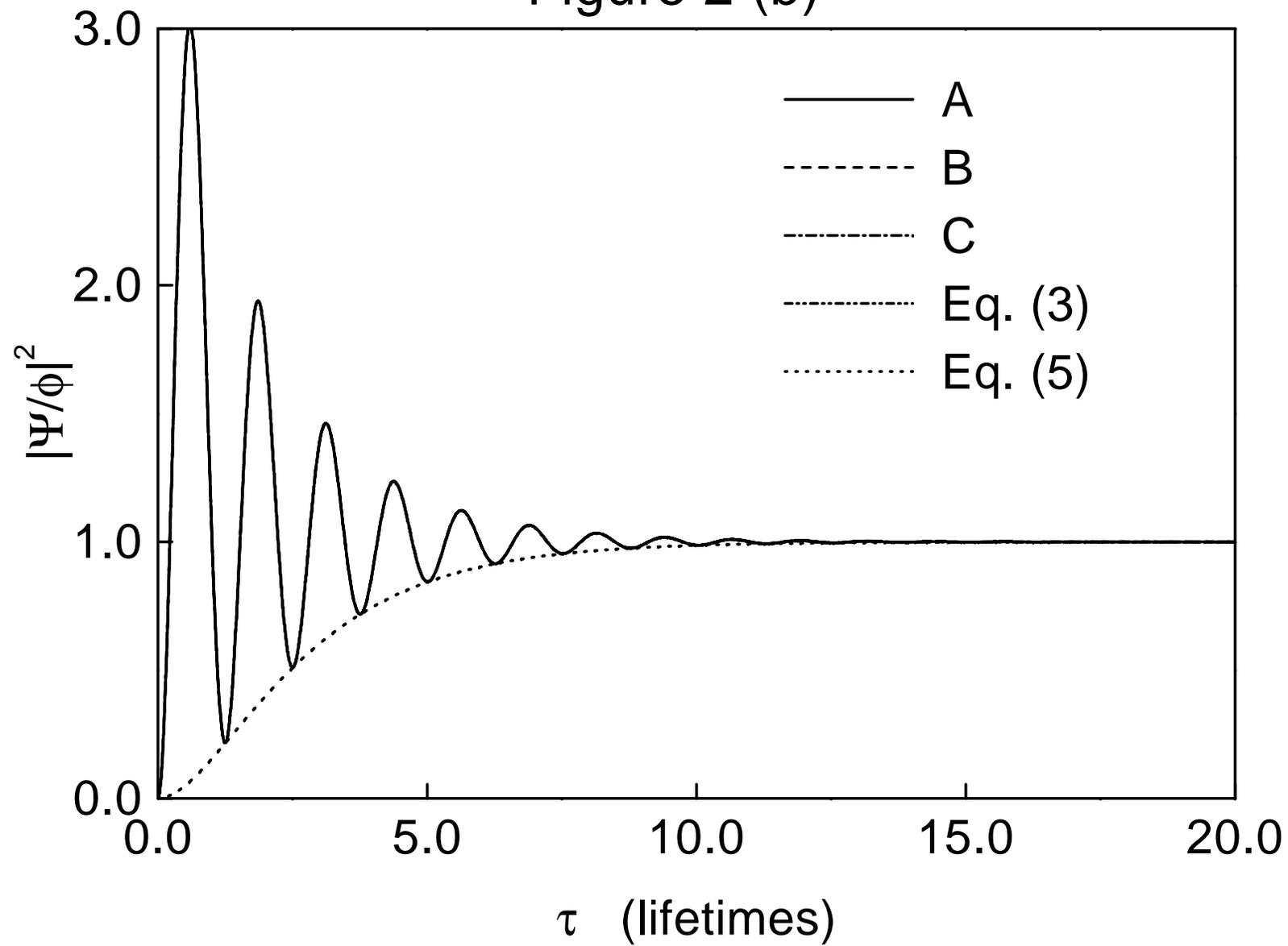
\begin{figure}[tbp]
\caption{ (a) The time evolution of $|\Psi |^2$ in the center of
the well at off-resonance incidence energy using Eq. (2) for systems A, B,
and C. (b) Also from Eq. (2), shows the time evolution of $|\Psi (%
\protect\tau )/\protect\phi |^2$ as a function of the time $\protect\tau$
given now in lifetime units; the curves of A, B and C are indistinguishable
among them. The calculation using the one-level formula, Eq. (3), is also
included in (b) for comparison and is indistinguishable from A, B, and C.
The lower envelope calculated from Eq. (5) is also shown. }
\label{fig2}
\end{figure}

\end{document}